\documentstyle[epsfig]{aprim10}
\input{epsf}
\newif\ifAMStwofonts
\ifoldfss
  \ifCUPmtlplainloaded \else
    \NewTextAlphabet{textbfit} {cmbxti10} {}
    \NewTextAlphabet{textbfss} {cmssbx10} {}
    \NewMathAlphabet{mathbfit} {cmbxti10} {} % for math mode
    \NewMathAlphabet{mathbfss} {cmssbx10} {} %  "   "    "
  \fi
  \ifAMStwofonts
    \ifCUPmtlplainloaded \else
      \NewSymbolFont{upmath} {eurm10}
      \NewSymbolFont{AMSa} {msam10}
      \NewMathSymbol{\upi}     {0}{upmath}{19}
      \NewMathSymbol{\umu}     {0}{upmath}{16}
      \NewMathSymbol{\upartial}{0}{upmath}{40}
      \NewMathSymbol{\leqslant}{3}{AMSa}{36}
      \NewMathSymbol{\geqslant}{3}{AMSa}{3E}

      \let\leq=\leqslant 
       
    \fi
  \fi
\fi % End of OFSS

\ifnfssone
  \newmathalphabet{\mathit}
  \addtoversion{normal}{\mathit}{cmr}{m}{it}
  \addtoversion{bold}{\mathit}{cmr}{bx}{it}
  \newmathalphabet{\mathbfit} % math mode version of \textbfit{..}
  \addtoversion{normal}{\mathbfit}{cmr}{bx}{it}
  \addtoversion{bold}{\mathbfit}{cmr}{bx}{it}
  \newmathalphabet{\mathbfss} % math mode version of \textbfss{..}
  \addtoversion{normal}{\mathbfss}{cmss}{bx}{n}
  \addtoversion{bold}{\mathbfss}{cmss}{bx}{n}
  \ifAMStwofonts
    \ifCUPmtlplainloaded \else
      \UseAMStwoboldmath
      \makeatletter
      \new@mathgroup\upmath@group
      \define@mathgroup\mv@normal\upmath@group{eur}{m}{n}
      \define@mathgroup\mv@bold\upmath@group{eur}{b}{n}
      \edef\UPM{\hexnumber\upmath@group}
      \new@mathgroup\amsa@group
      \define@mathgroup\mv@normal\amsa@group{msa}{m}{n}
      \define@mathgroup\mv@bold\amsa@group{msa}{m}{n}
      \edef\AMSa{\hexnumber\amsa@group}
      \makeatother
      \mathchardef\upi="0\UPM19
      \mathchardef\umu="0\UPM16
      \mathchardef\upartial="0\UPM40
      \mathchardef\leqslant="3\AMSa36
      \mathchardef\geqslant="3\AMSa3E

      \let\leq=\leqslant 

    \fi
  \fi
\fi % End of NFSS release 1

\ifnfsstwo
  \DeclareMathAlphabet{\mathbfit}{OT1}{cmr}{bx}{it}
  \SetMathAlphabet\mathbfit{bold}{OT1}{cmr}{bx}{it}
  \DeclareMathAlphabet{\mathbfss}{OT1}{cmss}{bx}{n}
  \SetMathAlphabet\mathbfss{bold}{OT1}{cmss}{bx}{n}
  \ifAMStwofonts
    \ifCUPmtlplainloaded \else
      \DeclareSymbolFont{UPM}{U}{eur}{m}{n}
      \SetSymbolFont{UPM}{bold}{U}{eur}{b}{n}
      \DeclareSymbolFont{AMSa}{U}{msa}{m}{n}
      \DeclareMathSymbol{\upi}{0}{UPM}{"19}
      \DeclareMathSymbol{\umu}{0}{UPM}{"16}
      \DeclareMathSymbol{\upartial}{0}{UPM}{"40}
      \DeclareMathSymbol{\leqslant}{3}{AMSa}{"36}
      \DeclareMathSymbol{\geqslant}{3}{AMSa}{"3E}

      \let\leq=\leqslant 

    \fi
  \fi
\fi % End of NFSS release 2

\ifCUPmtlplainloaded \else
  \ifAMStwofonts \else % If no AMS fonts
    \def\upi{\pi}
    \def\umu{\mu}
    \def\upartial{\partial}
  \fi
\fi

\title{Beta Viscose Prescription in Self-Gravitating Disks}

\author[Abbassi et al.]
       {S. Abbassi$^1$ and J. Ghanbari$^2$\\
        $^1$School of Physics, Damghan University of Basic Sciences,
Damghan, P.O.Box 36715-364, Iran\\
        $^2$Department of Physics, Ferdowsi University of Mashhad,
Mashhad, Iran}
\date{}

\pagerange{\pageref{firstpage}--\pageref{lastpage}}
\pubyear{2008}

\begin{document}

\maketitle

\label{firstpage}

\begin{abstract}
Duschl et al. (2000) have shown that the standard model for
geometrically thin accretion disks ($\alpha$-disks) leads to
inconsistency if self-gravity play a role. This problem arise from
parametrization of viscosity in terms of local sound velocity and
vertical disks scale hight. The $\beta$-viscosity prescription was
introduced by Duschl et al. (2000), which has been derived from
rotating shear flow experiment ($\nu=\beta \Omega R^2$). Following
the Duschl et al. (2000) suggestion for a $\beta$-prescription for
viscosity, we apply this model for a thin self-gravitating disk
around newborn stars. Our result is quite different with standard
alpha disks in the outer part of the disks where the self-gravity
becomes important. In the inner part of the disks, our solution
converged to the standard $\alpha$ disks. It has been presented that
for beta model, Toomre parameter is more than unity everywhere which
means that gravitational fragmentation can be occur everywhere. We
suggest that the kind of hydrodynamically driven viscosity,
$\beta$-model, can be used for modeling of accretion disks from
proto-stellar disks to AGN and galactic disks. It would be interest
to investigate ADAF-type solution for follow any effects by
$\beta$-viscosity model. An important property of the $\beta$-disk
is that they are viscously stable.
\end{abstract}


\section{Introduction}

Disk-like or flattened geometries are very common in astrophysics,
from large scale of spiral galaxies down to the small scales of
Saturn rings. The dynamics underlying the development of such
structures is determined by the propagation of density waves and the
important role of the disks self-gravity in their development has
been clearly recognized. In the standard thin disks model, the
effect of self-gravity is neglected, and only pressure supports the
vertical structure. In contrast, the theory of self-gravitating
accretion disks was less developed. From the observational point of
view, there are already some clues that the disk self-gravity could
be important both in the context of proto-stellar disks and in the
accretion disks around super-massive black holes in the AGN.

The study of self-gravity generally, is difficult and most authors
usually study the effects that relate to self-gravity either in the
vertical structure of the disk or in the radial direction. Usually
self-gravity occurs at large distances from the central objects and
mainly in the direction perpendicular to the plan of the disk. But
in the accretion disks around young stellar objects or
pre-main-sequence stars, self-gravity can be important in all parts
of the disk in both vertical and radial directions. Ghanbari \&
Abbassi (2004) introduced a model that shows self-gravity is an
important effect in the equilibrium structure of thick disks around
a compact object.

Viscous accretion through disks and the ensuing dissipation is known
to be a very efficient process of converting gravitational energy
into radiation. In particular accretion into the black holes allows
to librate a sizeable fraction of the accreted matter'Duschl et al.
(200. 00)076i8. nc s rest energy. Because detailed modeling of the
structure and evolutions of accretion disks depends on the viscosity
and its dependence on the physical parameters, choosing the best
viscosity model is important. There is a belief that the molecular
viscosity is inadequate to describe luminous accretion disks so that
some kind of turbulence viscosity is required. Most investigators
adopt the so-called $\alpha$-model introduce by Shakura \& Sanyev
(1972) and Shakura \& Sanyev (1973) that give the viscosity, $\nu$,
as the product of the pressure scale hight in the disk (h), the
velocity of sound, and a parameter $\alpha$ that contains all the
unknown physics. One interprets this as some kind of isotropic
turbulence viscosity $\nu=\nu_t=l_t v_t$ where $l_t$ is an length
scale and $v_t$ an characteristic velocity of a turbulence. One may
then write $\alpha=(\frac{v_t}{c_s})\cdot(\frac{l_t}{h})$. On
general physics grounds neither term in parenthesis can exceed unity
so that $\alpha\leq 1$. If initially $v_t>c_s$, shock waves would
result in strong damping and hence a return to a subsonic turbulence
velocity. The condition $l_t>h$ would require anisotropic turbulence
since the vertical length scales are limited by the disk's
thickness, which is comparable to h. The models for the structure
and evolution of accretion disks in close binary systems show that
$\alpha$-model leads to a result that reproduce the overall observed
behavior of the disks quite well. The $\alpha$-prescription is based
on turbulence viscosity but there is no physical evidence for this
as origin of turbulence.

\section{Condition for Self-gravity in Accretion Disk}

The main properties of accretion disks is that they are often thin.
This means that the typical length scale in the vertical direction,
the disk thickness h, is much smaller than the radial distance from
the central object. For example AGN disks are generally very thin,
with aspect ratio $h/r\approx0.001-0.01$. While proto-stellar disks
are comparably a bit thicker, with $h/r\approx0.1$. This has a
significant impact on the conditions under which the disk is
self-gravitating, and hence this intrinsic difference between the
AGN case and the proto-stellar case will be reflected in a
significant different behavior of  these two kind of system in the
self-gravitating regime.

One can estimate the importance of self-gravity by comparing the
respective contributions to the local gravitational accelerations in
the vertical and radial directions. The vertical gravitational
acceleration at the disk surface is $2\pi G\Sigma$ and
$GM_{*}h/r^3$, respectively, for the self-gravitating and the purely
Keplerian case. Self-gravitation is thus dominated in the vertical
direction when
$$
\frac{M_d}{M*}\sim\frac{\pi r^2 \Sigma}{M_*}>\frac{1}{2}\frac{h}{r}
$$

For typical proto-stellar and AGN disks with $h/r\sim0.001-0.1$ this
conditions translates into a condition for occurrence of vertical
self-gravitation of $M_d/M_*>0.0005-0.05$. Similar consideration
leads us a similar condition for occurrence of self-gravity in
radial direction. So when the mass of the disks is comparable with
the mass of central star self-gravity in all kind of thin disks,
from AGN to proto-stellar disks play an important role.

\section{Viscose Enigma in Accretion Disks}

The evolution of the disk and its time scale are governed by the
value of the viscosity parameter $\nu$. The viscose timescale
$\tau_{visc}$ is given by:

$$
\tau_{visc}=\frac{R^2}{\nu}
$$

It is not disputed that molecular viscosity is too small by many
orders of magnitude and leads in almost all relevant situations to
timescales surpassing the Hubble time. Some people believed that
some kind of turbulence viscosity can solve this problem. This
impasse was solved originally by Shakura\& Sanyev (1973). The
$\alpha$ prescription is based on the insight the molecularly
viscose accretion disks are pron to exceedingly large Reynolds
number, indicative of the onset of turbulence. It is worth noting
that widely used $\alpha$-prescription for turbulence viscosity dose
note invoke any particular mechanism at the origin of turbulence, it
is a simple parametrization which is tailored to yield turbulence
velocities that remain subsonic for $\alpha<1$.

Despite a number of successful applications of the $\alpha$
prescription (for instance in explaining the dwarf nova phenomenon),
this parametrization suffers from a number of inconveniences.

Duschl et al (2000) have shown that the standard model of thin
accretion disk based on $\alpha$ model lead to inconsistences if
self-gravity play an important role. This problem arises from the
parametrization of viscosity in terms of local sound velocity and
vertical disk scale hight.

For these multiple reasons it makes sense to investigate the basic
properties of accretion disks built with the alternate
$\beta$-prescription, which is observed in laboratory experiment
shear flows. Recently some laboratory experiments show that in a
self-gravitating flow Reynolds number is extremely high, it was
thought that hydrodynamically turbulence viscosity probably plays an
important role in the distribution of angular momentum in the
accretion disks (Richard \& Zahn 1999, Hure et al. 2001). Duschl et
al. (2000) have proposed a generalized accretion disks viscosity
prescription based on the hydrodynamically-driven turbulence at the
critical effective Reynolds number, the $\beta$-model, which is
applied to both self-gravitating and non-self-gravitating disks and
they have shown that it leads to standard $\alpha$-model.

\section{Comparing the $\alpha$ and $\beta$- viscosities}

The laboratory experiment have been set up to study of
hydrodynamical instabilities in differential rotating flows: a fluid
is sheared between two cylinders rotating at different speeds. Only
a few experiments have been run in the case where the angular
momentum increase inwards, as in a Keplerian accretion disks, but it
appears that the turbulence viscosity scales as:
$$
\nu_t\propto R^3\mid\frac{d\Omega}{dR}\mid
$$
where R is the distance from the rotation axis and $\Omega$ is
angular velocity (Richard \& Zahn 1999). In the Keplerian disks this
formula is equivalent to:
$$
\nu_{\beta}=\beta\Omega R^2=\beta R v_{\phi}
$$
which we shall call $\beta$-prescription.

One important property of the $\beta$-viscosity is that it depends
only on the radius in a Keplerian disk, and dose not involve local
physical conditions while the standard $\alpha$-prescription depends
on the local values of pressure scale hight $H$ and sound velocity
$c_s$:
$$
\nu_{\alpha}=\alpha c_s H
$$

Duschl et al. 2000 makes the sensible choice of equating the
parameter $\beta$ with the inverse the critical Reynolds number
\textbf{\textit{Re}}, which they assume to be the same as in plane
parallel shear flow (\textbf{\textit{Re}}$\approx 10^3$ and
therefore $\beta\approx 10^3$). However Hure et al. (2001) have
shown that more smaller value for $\beta$ can be obtained which is
derived from Couette-Taylor experiment. The only experimental data
available in the case where angular momentum increases outward (such
a Keplerian disk), are those obtained by Wendt (1933) and Taylor
(1936), in which the inner cylinder is at rest; for this one derive
$\beta=10^{-5}$.

The $\alpha$-prescription has been designed such that the Mach
number is smaller than unity for $\alpha\leq 1$. In contrast,
subsonic turbulence is not guaranteed with the $\beta$-prescription.
This restriction on the subsonic turbulence velocities is dictated
by the fact that supersonic turbulence would be highly dissipative
and therefore difficult to sustain. Moreover, the turbulence
viscosity presented here, $\beta$-prescription, has been measured in
a liquid, and its applications to compressible fluid can only be
justified in subsonic range.

\section{Discussion and Remarks}

The $\beta$-prescriptions is based on the assumption that the
effective Reynolds Number of the turbulence dose not fall bellow the
critical Reynolds number. We think that this prescription yields a
consistent model for fully self-gravitating disks where Duschl et
al. have been shown the standard $\alpha$ model leads a
inconsistency. Such $\beta$-disks model may be relevant for
proto-planetary disks and Galactic disks. In the case of
proto-planetary disk Duschl et al. have yield spectra that are
considerably flatter than those due to non-self-gravitating disks,
in better agreement with the observed spectra of these objects.

Self-gravitation which modifies the hydrodynamical equilibrium of
accretion disks  and affected all dynamical behavior and structure.
In standard disk model self-gravitation was ignored. But to reach a
more realistic picture of accretion disk, it should be take into
account. In the proto-planetary disks self-gravity should be
important in outer part of the disks where the disks mass is
comparable with central accretor mass. Disk in AGN are thought to be
relatively light in the sense that $\frac{M_d}{M_*}$ should not
exceed a few percent. In AGN disks self-gravity paly a role but at
large distance from central object (Shlosman \& Begelman 1987).

It may seems that using other form of viscosity is not an important
issue, because one should just change the mathematical form of the
equations. But this changes the governing equations of the system
that can effect the dynamical behaviors of the disks. This led us to
explore the self-gravitating disks using other viscose
prescriptions. However all viscose prescription have
phenomenological backgrounds rather than physical confirmed
backgrounds. $\beta$-viscose model have a good theoretical and
experimental background.

 Following the Duschl et al. (2000) suggestion for a
$\beta$-prescription for viscosity, Abbassi et al have applied this
model for a thin self-gravitating disk around newborn stars. Their
results is quite different with standard alpha disks in the outer
part of the disks where the self-gravity becomes important. In the
inner part of the disks their solution converged to a standard
$\alpha$ disks. They have shown that for $\beta$-disk Toomre
parameter, Q, is less than one in all radial distances, while is
more than one $\alpha$ model. As we know, when $Q<1$ the
gravitational fragmentation can be accrue. So gravitational
fragmentation can be accrue in all radial distances and it can be
used for planet formation in proto-planetary disks. They also show
that in a $\beta$ model accretion rate and radial flow is more than
the standard $\alpha$ model which means that this new prescription
is more effective than the old version. It can reproduce accretion
system in a less timescales.

Although the $\beta$-viscosity and the $\alpha$-viscosity may have
comparable magnitude, they have very different effects both on the
structure and on the stability steady Keplerian disks. It would be
interest to investigate ADAF-type solution for follow any effects by
$\beta$-viscosity model. An important property of $\beta$-disk is
that they are viscously stable. Also unlike the  $\alpha$-disks they
are thermally stable for ideal cooling processes, such as Thompson
scattering and free-free absorbtion (Hure et al. 2001).

In order to study to model and derive a realistic picture of a thin
self-gravitating disk, one must investigate the energy exchange of
the disks with environment. This require a mechanism to transfer the
thermal energy from the disk to the outside.

Both $\alpha$ and $\beta$ models are phenomenological prescriptions
for disk viscosity. In an actual model of viscosity, it is possible
to combine these two model and establish and exact description for
different regions of the disks. Also it can be prove by SPH
simulation of a self-gravitating flow.

\section*{Acknowledgment}
This research was funded under a grant from Damghan University of
Basic Sciences (DUBS).

\label{lastpage}

\clearpage

\end{document}